\documentclass{appolb}
\usepackage{graphicx}
\usepackage{amsmath}
\usepackage{amssymb}
\usepackage{lipsum}
\usepackage{lineno}
\usepackage{braket}
\usepackage[numbers]{natbib}

\makeatletter
\def\blfootnote{\xdef\@thefnmark{}\@footnotetext}
\makeatother

\begin{document}

\title{An Investigation of Charm Quark Jet Spectrum and Shape Modifications in Au+Au Collisions at $\sqrt{s_{\text{NN}}} = 200 \text{ GeV}$%
\thanks{Presented at Quark Matter 2022. This material is based upon work supported by the National Science Foundation under Grant No. 1913624.}%
}
\author{Diptanil Roy (\textit{For the STAR Collaboration}) \\ roydiptanil@gmail.com
\address{Rutgers University}
}
\maketitle
\begin{abstract}
Partons in heavy-ion collisions interact strongly with the Quark-Gluon Plasma (QGP), and hence have their energy and shower structure modified compared to those in vacuum. Theoretical calculations predict that the radiative energy loss, which is the dominant mode of energy loss for gluons and light quarks in the QGP, is suppressed for heavy quarks at low transverse momenta ($p_{\text{T}}$). At RHIC energies, lower energy jets closer to the charm quark mass are more accessible, and could provide key insight into the understanding of the mass dependence of parton energy loss. We report the first measurements of the $D^{0} (c\bar{u})$ meson tagged jet $p_{\text{T}}$ spectra and the $D^{0}$ meson radial profile in jets reconstructed from Au+Au collisions at $\sqrt{s_{\text{NN}}} = 200 \text{ GeV}$, collected by the STAR experiment.

\end{abstract}
  
\section{Introduction}

Relativistic heavy-ion collisions are able to produce Quark-Gluon Plasma (QGP), as predicted by Quantum Chromodynamics (QCD) \cite{STARQGP}. Internal probes involving hard scattering processes can be used to study the properties of the QGP medium. Jets, one of such probes, manifest as collimated clusters of final state particles in the detector. The partons which give rise to these jets lose energy to the QGP medium, either through elastic collisions, or through induced gluon \textit{bremsstrahlung} - a phenomenon known as jet quenching \cite{ReviewJets}. The effects of jet quenching can be seen in measurements of inclusive jets yield suppression \cite{InclusiveJetRCPSTAR} and modifications to the jet structure \cite{CMSJetShape}. A study of heavy flavor tagged jets can shed light on the mass and flavor dependence of the  parton energy loss and jet structure modifications. The dead-cone effect \cite{DeadConeTheory}, as predicted by the QCD, has been measured for charm quarks in \textit{pp} collisions at the LHC \cite{DeadCone}, but remains elusive in heavy-ion collisions. Heavy flavor jets at the LHC have also yet to reveal significant differences to their inclusive counterparts \cite{bjetQuenchingCMS, CMSD0JetPaper}, possibly due to having energies much higher than the parton mass. Such studies at the RHIC energies, where lower energy jets are produced, could be the key to better understand the parton mass dependence of the energy loss. This proceeding will focus on the first measurements of $D^{0} (\bar{D}^{0})$ meson tagged jet transverse momentum ($p_{\text{T}}$) spectra and the $D^{0} (\bar{D}^{0})$ meson radial profile in tagged jets from Au+Au collisions at $\sqrt{s_{\text{NN}}} = 200 \text{ GeV}$.
\section{Analysis Setup}
This work uses Minimum Bias (MB) triggered Au+Au events at $\sqrt{s_{\text{NN}}}$ = 200 GeV, collected in 2014 by the STAR detector \cite{STAROverview} at RHIC. Events and tracks, which pass standard quality cuts at STAR \cite{STARDijetPaper}, are chosen within the pseudorapidity acceptance of $|\eta| < 1$. This analysis is done in three centrality bins: 0-10\% (central), 10-40\% (mid-central), and \mbox{40-80\% (peripheral)}.
$D^{0} (\bar{D}^{0})$ mesons are reconstructed via the hadronic decay channel $D^0 \rightarrow K^- + \pi^+$(and its charge conjugate) with a branching ratio of 3.89 \% \cite{PDG}. Several topological selections based on the decay geometry of $D^{0} (\bar{D}^{0})$ are applied to supress the combinatorial $K\pi$ pairs in an event, thanks to the excellent track pointing resolution provided by the Heavy Flavor Tracker (HFT) \cite{HFT}. A more thorough discussion on the selection criteria for the $D^{0} (\bar{D}^{0})$ candidates is available in Ref. \cite{STARD0PaperAuAu}.

Full jets are reconstructed from TPC tracks and electromagnetic calorimeter (ECAL) towers with $p_{\rm T} > 0.2$ $\rm GeV/\textit{c}$, and transverse energy \mbox{$E_{\rm T} > 0.2$ GeV} respectively. Jets are found using the anti-$k_{\rm T}$ clustering algorithm available in the FastJet package \cite{FASTJET}, with a radius parameter of $R = 0.4$ in the $\eta-\phi$ space, and are selected in the pseudorapidity range $|\eta_{\rm jet}| < 1 - R$. The $K$ and $\pi$ daughter tracks are replaced with the corresponding $D^{0} (\bar{D}^{0})$ candidate before the jets are reconstructed. A jet area based background subtraction is applied to remove the average background contribution to the jet energy \cite{JetAreaBackground}. Jets with a $D^{0} (\bar{D}^{0})$ constituent of $p_{\rm T, D^{0}} \in (5, 10) \text{ GeV}/\textit{c}$ are considered as a $D^{0}$ tagged jet for this analysis.

\section{$D^{0} (\bar{D}^{0})$ Jet Spectra and Shape Modifications}
\begin{figure}[t]
    \centering
    \includegraphics[width=0.55\textwidth]{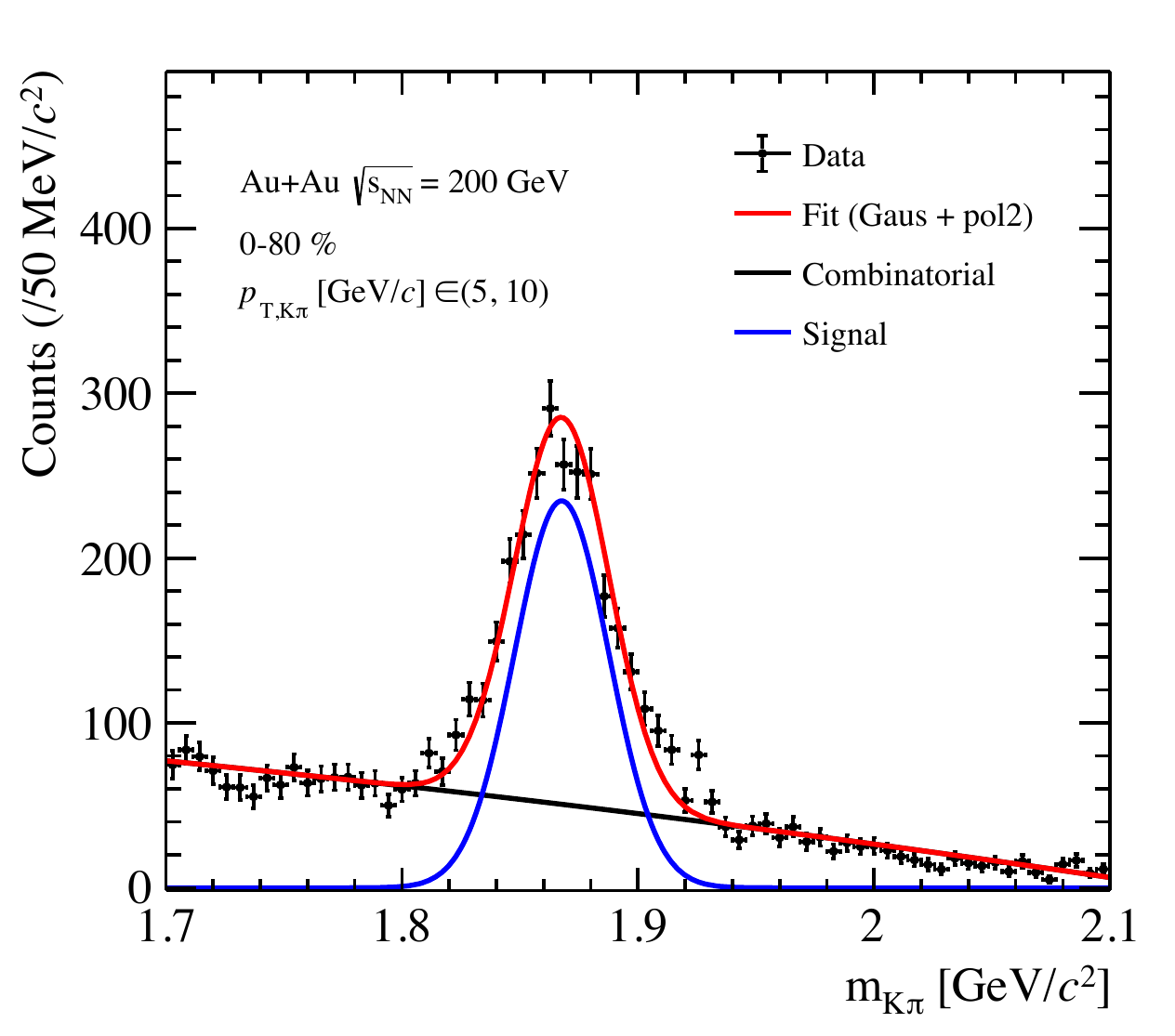}
    \caption{The invariant mass distribution of $K\pi$ pairs with $p_{\rm T} \in (5, 10) \text{ GeV}/\textit{c}$. The unlike sign $K\pi$ pair distribution (\textit{black}) is fit with a Gaussian plus second-order polynomial function (\textit{red}) to estimate the $D^{0} (\bar{D}^{0})$ yield. The signal after the removal of the background (\textit{blue}) is also shown.}
    \label{fig:Invariant mass}
\end{figure}

To extract the raw yield of $D^{0} (\bar{D}^{0})$ mesons, a method called $_s$\( \mathcal{P} \)\textit{lot} \cite{SPlot} is used. $_s$\( \mathcal{P} \)\textit{lot} calculates per event weights, called sWeights, from an unbinned likelihood fit to the $D^{0} (\bar{D}^{0})$ invariant mass distribution. The weight classifies how likely it is for a $D^{0} (\bar{D}^{0})$ candidate to be a true $D^{0} (\bar{D}^{0})$. \mbox{Figure \ref{fig:Invariant mass}} shows the invariant mass distribution of $K\pi$ pairs in the $p_{\rm T}$ region of \mbox{5–10 $\rm GeV/c$} for 0–80\% MB events. The raw $D^{0}$ jet distributions are obtained by weighing each candidate with the corresponding sWeight.
The invariant yield of $D^{0} (\bar{D}^{0})$ tagged jets is represented by the formula:
\begin{equation}
    \frac{d^2N_{\rm jet}}{2\pi \: N_{\rm evt} \: p_{\rm T,jet} \: dp_{\rm T,jet} \: d\eta} = \frac{1}{\rm B.R.} \times \frac{N^{\rm raw}_{\rm jet}}{2\pi \: N_{\rm evt} \: p_{\rm T,jet} \: \Delta p_{\rm T,jet} \: \Delta\eta} \times \frac{1}{\epsilon_{\rm corr}}
\end{equation}
where $\rm B.R.$ is the $D^{0} \rightarrow K^{-}\pi^{+}$ decay branching ratio ($3.89 \pm 0.04 \%$), $N^{\rm raw}_{\rm jet}$ is the reconstructed $D^{0} (\bar{D}^{0})$ tagged jet raw counts, and $N_{\rm evt}$ is the total number of events used in this analysis. The raw yields are corrected for the tracking efficiency and acceptances of the TPC and HFT, topological cut efficiency, particle identification efficiency, and finite vertex resolution based on the correction factors derived in Ref. \cite{STARD0PaperAuAu}, and the total correction factor is $\epsilon_{\rm corr}$. The nuclear modification factor $R_{\rm CP}$ is defined as the
ratio of $\left\langle\rm N_{\rm coll}\right\rangle$–normalized yields between central and peripheral collisions, where $\left\langle\rm N_{\rm coll}\right\rangle$ is the average number of binary collisions for a centrality class.

The radial distribution of $D^{0} (\bar{D}^{0})$ mesons in tagged jets is defined by the formula:
\begin{equation}
    \frac{1}{N_{\rm jet}}\frac{dN_{\rm jet}}{d\rm r} =  \frac{1}{N_{\rm jet}}\frac{N_{\rm jet}|_{\Delta \rm r}}{\Delta \rm r}
\end{equation}
where $\rm r = \sqrt{(\eta_{\rm jet} - \eta_{D^{0}})^2 + (\phi_{\rm jet} - \phi_{D^{0}})^2}$ is the distance of the $D^{0}$($\eta_{D^{0}}, \phi_{D^{0}}$) from the jet axis ($\eta_{\rm jet}, \phi_{\rm jet}$) in the $\eta-\phi$ plane, and $N_{\rm jet}|_{\Delta \rm r}$ is the number of jets with $D^{0} (\bar{D}^{0})$ mesons in the $\Delta \rm r$ interval.


A Bayesian unfolding procedure \cite{Unfolding} is used to account for the detector inefficiencies in jet reconstruction. A $D^{0} (\bar{D}^{0})$-enriched sample of $pp$ events at $\sqrt{s} = 200 \text{ GeV}$ is generated using PYTHIA v8.303, with the `Detroit' tune \cite{STARTune}, and propagated through the STAR detector simulation using the GEANT3 \cite{GEANT3} package. The charm quark spectrum based on FONLL \cite{FONLL} is used as a prior in the unfolding procedure. The charm quark fragmentation function is modeled using PYTHIA, and a systematic study of its variation is in the works. Observables with an asterisk(*), found later in this proceeding, are corrected with the PYTHIA fragmentation function.
The fluctuation due to the heavy-ion background is estimated by embedding one  `single-particle' jet in each MB Au+Au event, and then matching each embedded jet with a reconstructed jet containing the tagged `single-particle' \cite{SPJet_JetQuenching}. The quantity $\Delta p_{\rm T, SP jet} = p_{\rm T, SP jet}^{\rm det} - p_{\rm T, SP jet}^{\rm part}$ models this fluctuation. The superscript `part' refers to particle-level jets, and `det' refers to detector-level jets.
For the $D^{0}$ meson radial profile, the aforementioned Bayesian unfolding procedure is used to simultaneously correct $N_{\rm jet}$ as a function of $p_{\rm T,jet}$ and $\Delta \rm r$. 



The systematic uncertainties in the reported observables are dominated by the following contributions: a) differences in the invariant yield of $D^{0}$ mesons calculated using the $_s$\( \mathcal{P} \)\textit{lot} method, and a like-sign background subtraction method, and b) systematic uncertainty in $D^{0} (\bar{D}^{0})$ reconstruction efficiency taken from Ref. \cite{STARD0PaperAuAu}. Systematic variations related to the unfolding procedure are estimated by varying the following: a) the prior from FONLL to the $D^{0}$ tagged jet distribution generated by PYTHIA, and b) the regularisation parameter.

\begin{figure}[h]
    \centering
    \includegraphics[width=0.49\textwidth]{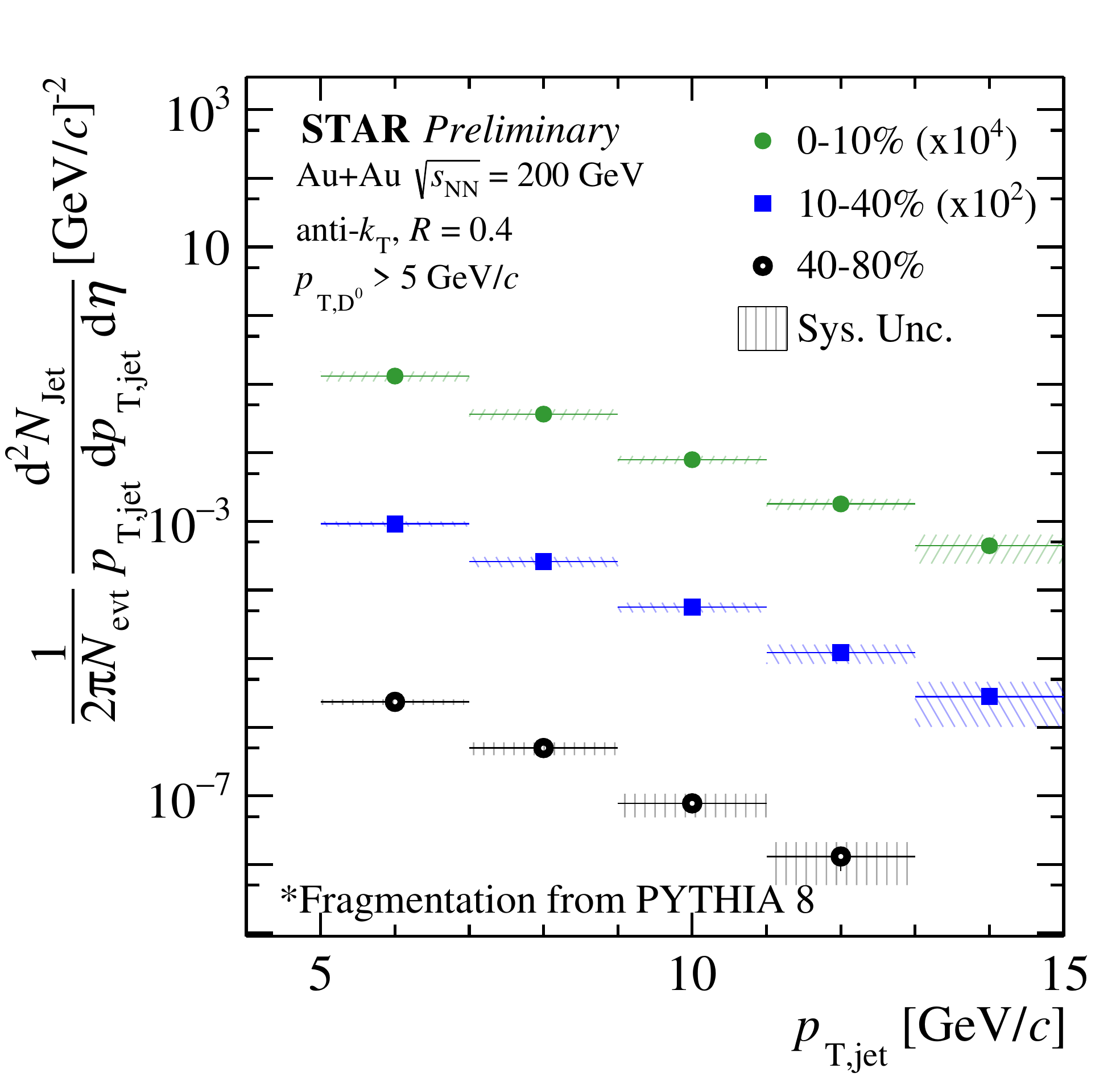}
    \includegraphics[width=0.49\textwidth]{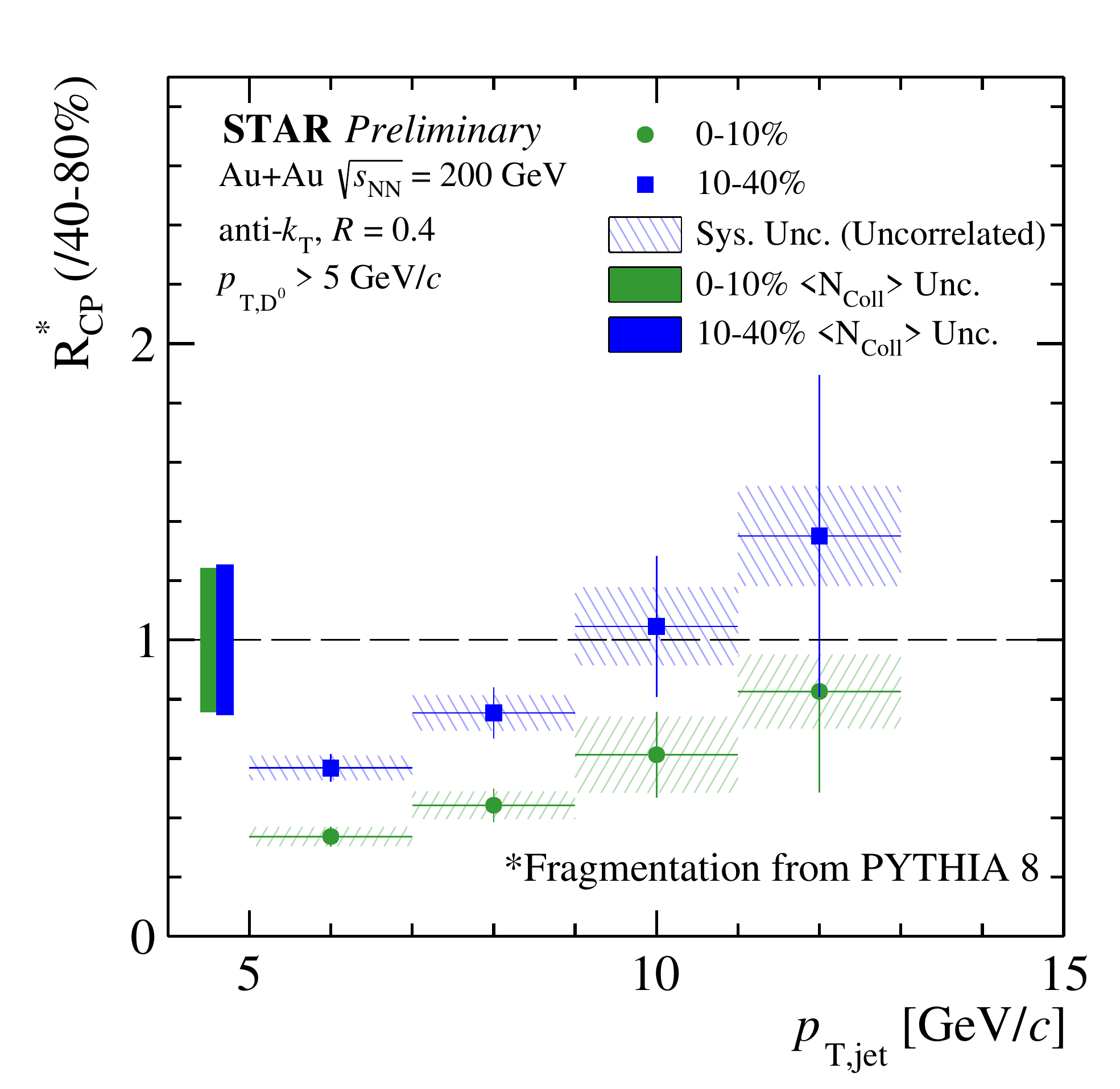}
    \caption{\textbf{Left}: $D^{0} (\bar{D}^{0})$ tagged jet $p_{\rm T}$ spectra with $p_{\rm T, D^{0}} \in (5, 10) \text{ GeV}/\textit{c}$ in different centrality classes; \textbf{Right}: Nuclear modification factor $R_{\rm CP}^*$ for $D^{0}$ jets.}
    \label{fig:Spectrum}
\end{figure}

The efficiency-corrected invariant yield of $D^{0} (\bar{D}^{0})$ jets is shown in the left panel of Fig. \ref{fig:Spectrum} for $p_{\rm T, D^{0}} \in (5, 10) \text{ GeV}/\textit{c}$, as a function of $p_{\rm T,jet}$ in \mbox{0-10\%}, 10-40\%, and 40-80\% Au+Au collisions. The spectra in the first two centrality bins are scaled by arbitrary factors for better visibility. The nuclear modification factor $R_{\rm CP}^*$ for the central and mid-central Au+Au collisions are shown in the right panel of Fig. \ref{fig:Spectrum}, with the peripheral centrality bin as the reference. The bands (blue and green) at unity are uncertainties associated with $\left\langle\rm N_{\rm coll}\right\rangle$. The $D^{0}$ jet $R_{\rm CP}^*$ shows a stronger suppression in central collisions than in mid-central collisions at low $p_{\rm T, jet}$. $R_{\rm CP}^*$ also shows an increasing trend with $p_{\rm T, jet}$ for both centrality bins. This trend is qualitatively different from the $R_{\rm CP}$ measured for inclusive jets at \mbox{RHIC \cite{InclusiveJetRCPSTAR}.} 

\begin{figure}[h]
    \centering
    \includegraphics[width=0.49\textwidth]{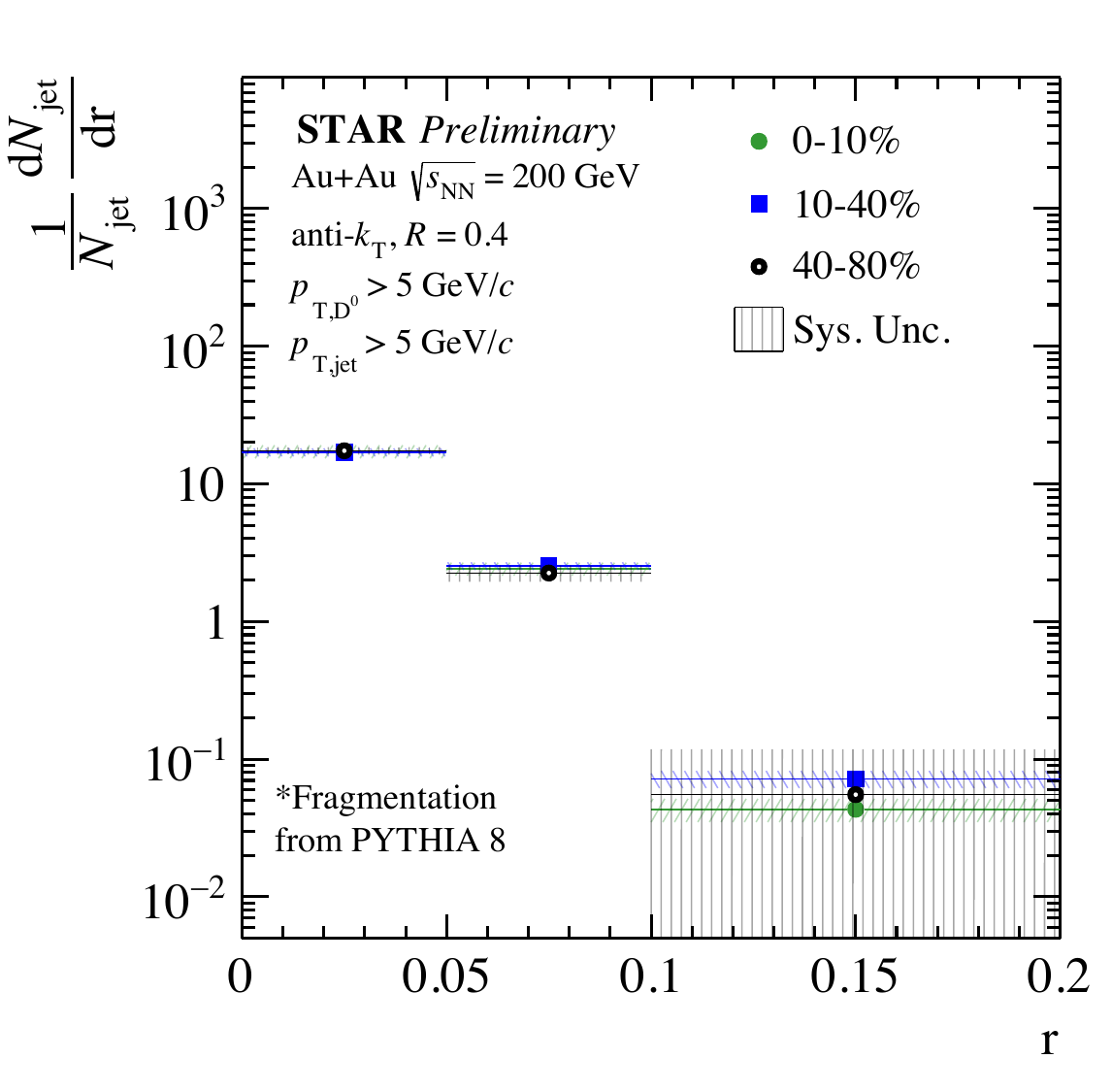}
    \includegraphics[width=0.49\textwidth]{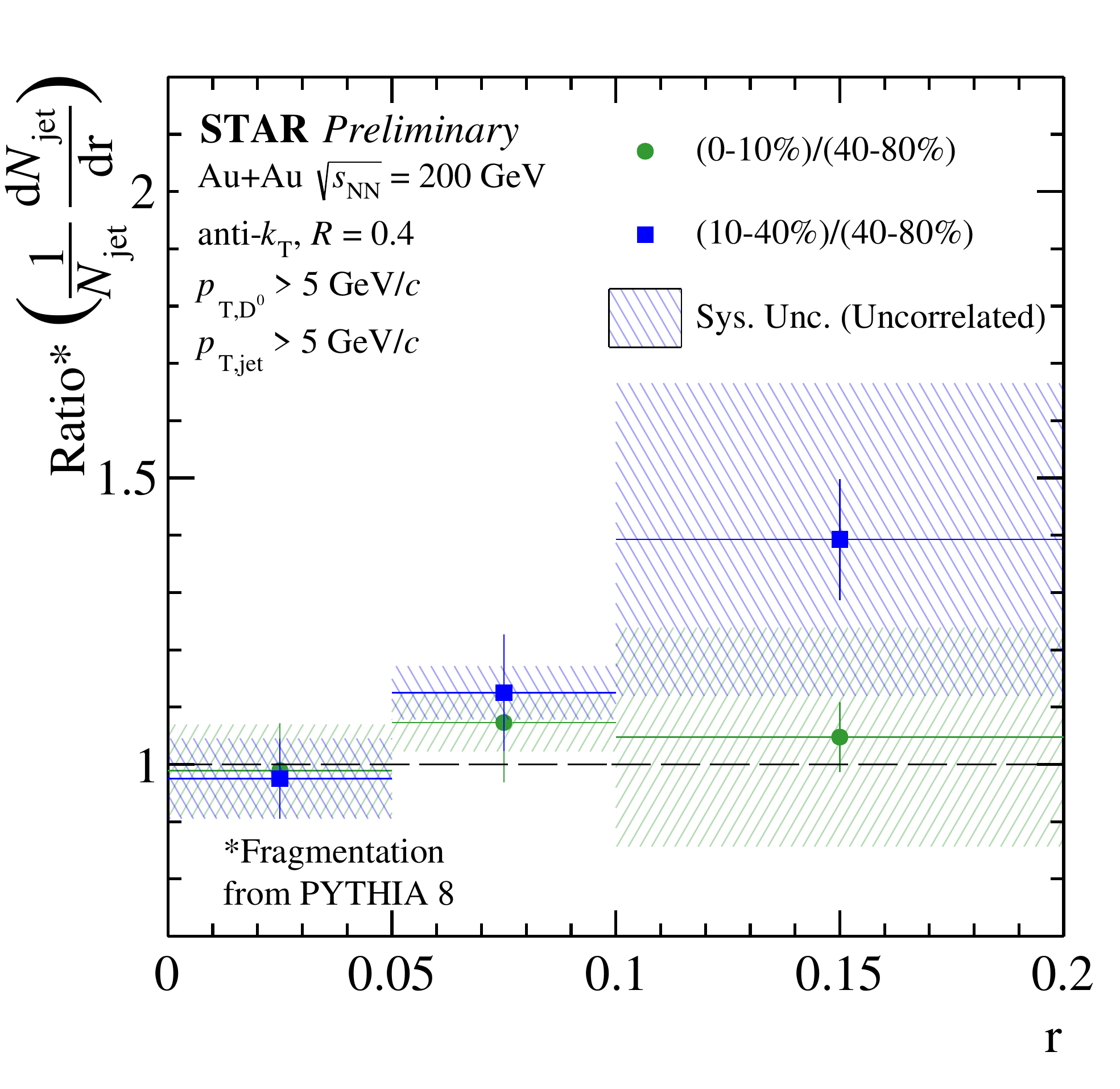}
    \caption{\textbf{Left}: $D^{0}$ radial profile in $D^{0} (\bar{D}^{0})$ tagged jets with $p_{\rm T, D^{0}} \in (5,10) \text{ GeV}/\textit{c}$ in different centrality classes; \textbf{Right}: Ratio of $D^{0}$ radial profiles for central and mid-central events with respect to $D^{0}$ radial profile for peripheral events.}
    \label{fig:RadialProfile}
\end{figure}

The radial profile for $D^{0} (\bar{D}^{0})$ mesons with $p_{\rm T, D^{0}} \in (5, 10) \text{ GeV}/\textit{c}$ in the tagged jets is shown as a function of the distance from the jet \mbox{axis ($\rm r$)} in \mbox{0-10\%}, 10-40\%, and 40-80\% Au+Au collisions in the left panel of \mbox{Fig. \ref{fig:RadialProfile}.} The ratios of the radial profiles for the central and mid-central events to peripheral events, shown in the right panel of Fig. \ref{fig:RadialProfile}, are found to be consistent with unity within the uncertainties. The large uncertainties are dominated by the limited statistics in the peripheral centrality bin.

\section{Discussion}
In this proceeding, the first measurements of $D^{0}$ meson tagged jet $p_{\text{T}}$ spectra and $D^{0}$ meson radial profile are reported for $p_{\rm T, D^{0}} \in (5, 10) \text{ GeV}/\textit{c}$ in Au+Au collisions at $\sqrt{s_{\text{NN}}} = 200 \text{ GeV}$. The $D^{0}$ $p_{\rm T, jet}$ spectra are found to be suppressed for central and mid-central collisions at low $p_{\rm T, jet}$ with the nuclear modification factor showing an increasing trend with $p_{\rm T, jet}$. This trend is qualitatively different from the inclusive jet measurements at RHIC. The radial profile of $D^{0} (\bar{D}^{0})$ in its tagged jets is found to be consistent for different centralities. Further studies are ongoing to extend our measurements to lower $p_{\rm T, D^{0}}$ and $p_{\rm T, jet}$ allowing us to get even closer to the charm quark mass. These measurements can help constrain theoretical models on parton flavor and mass dependencies of jet energy loss.

\bibliography{QM_DiptanilRoy} 
\bibliographystyle{unsrt-notitle}



\end{document}